\documentclass[apj]{emulateapj}

\shorttitle{Using dust to map the ISM in high-redshift galaxies}
\shortauthors{Eales et al.}
\submitted{submitted to ApJ}
\usepackage{graphicx}
\begin{document}


\title{Can dust emission be used to map the interstellar medium
in high-redshift galaxies?
Results from the Herschel
Reference Survey}


\author{Stephen Eales\altaffilmark{1},
Matthew W. L. Smith\altaffilmark{1},
Robbie Auld\altaffilmark{1},
Maarten Baes\altaffilmark{2},
George J. Bendo\altaffilmark{3},
Simone Bianchi\altaffilmark{4},
Alessandro Boselli\altaffilmark{5},
Laure Ciesla\altaffilmark{5},
David Clements\altaffilmark{6},
Asantha Cooray\altaffilmark{7},
Luca Cortese\altaffilmark{8},
Jon Davies\altaffilmark{1},
Ilse De Looze\altaffilmark{2},
Maud Galametz\altaffilmark{9},
Walter Gear\altaffilmark{1},
Gianfranco Gentile\altaffilmark{2},
Haley Gomez\altaffilmark{1},
Jacopo Fritz\altaffilmark{2},
Tom Hughes\altaffilmark{10},
Suzanne Madden\altaffilmark{11},
Laura Magrini\altaffilmark{4},
Michael Pohlen\altaffilmark{1}, 
Luigi Spinoglio\altaffilmark{12},
Joris Verstappen\altaffilmark{2},
Catherine Vlahakis\altaffilmark{13},
Christine D. Wilson\altaffilmark{14}
}

\altaffiltext{1}{School of Physics and Astronomy,
Cardiff University, Queens Buildings, The Parade,
Cardiff CF24 3AA, UK}
\altaffiltext{2}{Sterrenkundig Observatorium, Universiteit Gent, Krijgslaan
281 S9, B-9000 Gent, Belgium}
\altaffiltext{3}{UK ALMA Regional Centre Node, Jodrell Bank Centre for Astrophysics,
School of Physics and Astronomy, University of Manchester, Oxford
Road, Manchester M13 9PL, UK}
\altaffiltext{4}{INAF-Osservatorio Astrofisico di Arcetri, Largo E. Fermi 5, 50125, Firenze, Italy}
\altaffiltext{5}{Laboratoire d'Astrophysique de Marseilles, UMR6110 CNRS,
38 rue F. Joliot-Curie, F-1338 Marseilles, France}
\altaffiltext{6}{Astrophysics Group, 
Imperial College, Blackett Lab, Prince Consort Road, London SW7 2AZ, UK}
\altaffiltext{7}{Dept. of Physics \& Astronomy, Univ. of California, Irvine, 
CA 92697, USA}
\altaffiltext{8}{European Southern Observatory, Karl-Schwarzschild-Strasse
2 D-85748, Garching bei Munchen, Germany}
\altaffiltext{9}{Institute of Astronomy, University of Cambridge, Madingley Road, Cambridge CB3 0HA, United Kingdom}
\altaffiltext{10}{Kavli Institute for Astronomy \& Astrophysics,
Peking University, Beijing 100871, China}
\altaffiltext{11}{Laboratoire AIM, CEA/DSM - CNRS - Universit ́e Paris Diderot, Irfu/Service d’Astrophysique, 91191 Gif sur Yvette, France}
\altaffiltext{12}{Istituto di Fisica dello Spazio Interplanetario,
Istituto Nazionale di Astrofisica,
Via Fosso del Cavaliere 100,
I-00133 Roma,
Italy}
\altaffiltext{13}{Joint ALMA Office, Alonso de Cordova 3107, Vitacura, and
Departamento de Astronomia, Universidad de Chile, Casilla 36-D, Santiago, Chile}
\altaffiltext{14}{Department of Physics \& Astronomy, McMaster University, Hamilton, Ontario L8S 4M1, Canada}



\begin{abstract}

It has often been suggested that an alternative to the
standard CO/21-cm method for estimating the mass of the interstellar
medium (ISM) in a galaxy might be to estimate
the mass of the ISM from the continuum dust emission. In this paper,
we investigate the potential of this technique using
Herschel
observations of ten galaxies in the Herschel Reference Survey and in the Herschel Virgo Cluster Survey.
We show that the emission detected by Herschel is mostly from dust that
has a temperature and emissivity index similar to that of dust
in the local ISM in our galaxy, with the temperature generally increasing
towards the centre of each galaxy. 
We calibrate the dust method using the CO and 21-cm observations to provide
an independent estimate of the mass of hydrogen in each galaxy, solving the
problem of the uncertain `X factor' for the molecular
gas by minimizing the dispersion
in the ratio of the masses estimated using the two methods.
With the calibration for the dust method and the estimate of
the X-factor produced in this way, the
dispersion in the ratio
of the two gas masses is 30\%, which 
gives an upper limit on the fundamental accuracy of the dust method.
The calibration we obtain for the dust method is very similar
to an independent Herschel measurement for M31 and to the calibration for
the Milky Way from Planck measurements.

\end{abstract}


\keywords{galaxies: ISM --- submillimetre}



\section{Introduction}

In order to understand the physics and evolution of galaxies, accurate 
measurements of their gas content are absolutely crucial: 
stars produce most of the energy output
of galaxies and they form out of this reservoir of gas.
Unfortunately there are problems with all the
techniques
for observing the different phases of the interstellar medium (ISM). Probably the
most reliable technique is to use the 21-cm line 
to map the
atomic phase, but 
if the gas becomes optically thick the linear
relationship between the line brightness and
the column density of gas breaks down (Braun et al. 2009).
The mass of the molecular phase is usually estimated
from the 1-0 line of the tracer CO molecule, but the constant of proportionality
between the two - the `X-factor' - is notoriously uncertain
(Bell et al. 2007), and there is evidence that
it depends on metallicity (Wilson 1995; Israel 2005; Boselli
et al. 2002) and possibly on other factors
(Israel 1997). An additional problem with the
standard CO/21-cm method is that recent observations with
Fermi and Planck imply that a significant
fraction of the gas in the Galaxy
consists of `dark gas' traced by neither line (Abdo et al.
2010; Planck Collaboration 2011a).
There is also the practical problem that with current telescopes it
is difficult to detect either line from galaxies
at $z > 0.1$. 

As early as the mid-eighties (Hilderbrand 1983) 
it was suggested
that one way to estimate the mass of the ISM in a galaxy might be from the
optical depth of the submillimetre
continuum emission from the dust; dust grains are robust and found in all phases
of the ISM and the continuum dust emission is generally optically thin. 
More recent attempts to use the dust emission to infer the gas distribution
are described in Boselli et al. (2002) and in Guelin et al. (1993, 1995).
The reason why this method is of topical
interest is that the
Herschel Space Observatory (Pilbratt et al. 2010) is in the 
process of measuring the continuum dust
emission from hundreds of thousands of galaxies, seen over 10 billion years of cosmic
history (Eales et al. 2010a; Oliver et al. 2011), and it will never be practical
to estimate the mass of the ISM in so many galaxies using the
standard 21-cm/CO method.
Although ALMA will improve this for CO observations,
it will still not be feasible to measure
the strength of the CO line in thousands of
high-redshift galaxies.
A recent Herschel result has given additional encouragement
that this method may be a useful one, since Corbelli et al. (2011) have shown that in
a sample of galaxies in the Virgo cluster dust mass is more tightly correlated to
the total gas mass than to the masses of molecular or atomic gas separately.

To apply the dust method, of course, it is
necessary to know the temperature of the dust, but 
this is a practical problem with the method rather than a fundamental
one, and in principle it can be solved with accurate measurements
of the continuum emission at enough far-infrared wavelengths
to cover the peak of the emission.
An additional practical problem is that there is evidence
that the ratio of dust emission
to gas mass depends on the metallicity of the gas (Lisefeld \& Ferrara
1998; James et al. 2002; Draine et al. 2007), although this is
obviously also a problem with the alternative method, which uses the CO line to
trace the molecular phase. 

One way to try to calibrate the relationship between the submillimetre optical-depth
and the mass of the ISM is
to do it in two steps: first obtain a relationship between the optical
depth and
the mass of dust; then obtain a relationship between the mass of dust and the mass
of the ISM.
The problem with this approach is that the uncertain radiative efficiency
of dust grains 
\citep{draine} and the lack of a reliable independent method of measuring the
gas-to-dust ratio in galaxies
mean
that there are difficulties in both steps (Hildebrand 1983).
In this paper we adopt the more direct approach of ignoring the
properties of the dust grains and calibrating the relationship
between gas mass and dust optical depth directly. 
Recent Planck results suggest this is a promising method. The Planck
team finds the relationship between submillimetre optical
depth and gas column density is independent of Galactic radius and
the same in both the atomic and molecular phases (Planck Collaboration
2011b,c).

In this paper, we use Herschel observations with SPIRE (Griffin et al.
2010) and PACS (Poglitsch et al. 2010)
of ten nearby galaxies
taken from the Herschel Reference Survey (Boselli et al. 2010) and
the Herschel Virgo Cluster Survey (Davies et al. 2010)
to estimate the relationship
between the dust optical-depth and the
mass of gas in each galaxy.

\section{The Method}

The Planck team (Planck Collaboration 2011b)
has recently used Planck
observations of the continuum emission from the
dust to examine the relationship
between the optical depth of continuum emission
from dust
and the column density of gas as a function
of Galactic radius. Using an X-factor for the molecular phase
of $\rm 1.8 \times 10^{20} cm^{-2} (K km\ s^{-1})^{-1}$, which is
consistent with recent studies of the diffuse gamma-ray emission
in the Galactic plane (Abdo et al. 2010; Ackermann et al. 2011),
the Planck team found the following relationship at the solar
circle:

\medskip
\begin{equation}
\tau  = 1.1\times10^{-25} ({\lambda \over 250{\rm \mu m}})^{-1.8} 
N_H 
\end{equation}
\medskip

\noindent in which 
$\tau$ is the optical depth at wavelength $\lambda$ and 
$N_H$ is the column density of hydrogen
measured in atoms
cm$^{-2}$. 
The team found that this relationship is independent of Galactic
radius and is
the same in the molecular and atomic
phases, although this latter conclusion depends on the value
of the X-factor being correct.
The relationship above is very similar to the result from the
Planck and COBE
observations of dust at high Galactic latitude (Planck collaboration
2011c; Boulanger et al. 1996). The Planck team found a 
value of the dust-emissivity index, $\beta$, of 1.8, which is independent
of Galactic radius, and they showed that the temperature of dust in the atomic phase 
falls
with Galactic radius, with a value of 17.6 K at the solar circle
(Planck Collaboration 2011b).
The relationship above can be translated (Appendix 1)
into the following relationship between mass of hydrogen, $M_H$, and
monochromatic submillimetre luminosity ($L_{\nu}$):

\smallskip

\begin{equation}
M_{H} = \eta_c \times {1.52 \times 10^2 \times L_{\nu} \over
B_{\nu}(T_d) \times ({\lambda \over 250})^{-1.8} } 
\end{equation}
\smallskip

\noindent 
We have put everything in equation (2) in S.I. units except wavelength ($\lambda$), which
is measured in microns. 
Luminosity is measured in Watts Hz$^{-1}$ sr$^{-1}$.
We have introduced a constant, $\eta_c$, into
the equation. This is equal to
one for the Milky Way. Our objective in this paper is to measure
$\eta_c$ for
external galaxies.

\begin{deluxetable*}{llllllllll}
\tabletypesize{\scriptsize}
\tablecaption{Sample}
\tablewidth{0pt}
\tablehead{
\colhead{Galaxy} & \colhead{D/Mpc} & \colhead{12 + log(O/H)} &
\colhead{PACS} & \colhead{$N$} & \colhead{$<\beta>$} & \colhead{$<T>$} &
\colhead{$r_s$ ($\beta$, $r$)} & \colhead{$r_s$ ($T$,$r$)}
}
\startdata
NGC4192 (M98) & 16.8 &  8.76$\pm$0.08 & Y & 29 & 2.11$\pm$0.04 & 15.5$\pm$0.2 & 0.21 & -0.56
\\
NGC 4254 (M99) & 16.8 & 8.71$\pm$0.13 & Y & 44 & 2.20$\pm$0.02 & 17.6$\pm$0.3 & -0.10 & -0.48
\\
NGC 4321 (M100) & 16.8 & 8.75$\pm$0.05 & Y & 53 & 2.56$\pm$0.04 & 15.5$\pm$0.2 & 0.39 & -0.63 \\
NGC 4402 & 16.8 & 8.67$\pm$0.02 & Y & 10 & 2.17$\pm$0.09 & 19.7$\pm$0.7 & -0.23 & 0.20 \\
NGC 4419 & 16.8 & ... & N & 7 & 2.20$\pm$0.08 & 21.5$\pm$0.6 & 0.02 & -0.10 \\
NGC 4535 & 16.8 & 8.75$\pm$0.05 & Y & 35 & 2.24$\pm$0.02 & 16.1$\pm$0.3 & 0.43 & -0.65 \\
NGC 4536 & 16.8 &  8.71$\pm$0.08 & N & 42 & 1.43$\pm$0.04 & 24.9$\pm$0.1 & -0.73 & -0.22 \\
NGC 4569 (M90) & 16.8 &  ... & N & 19 & 2.29$\pm$0.05 & 19.5$\pm$0.3 & 0.39 & -0.61 \\
NGC 4579 (M58) & 16.8 & ... & Y & 26 & 2.26$\pm$0.06 & 17.4$\pm$0.3 & 0.31 & -0.78  \\
NGC 4689 & 16.8 & 8.66$\pm$0.05 & N & 16 & 2.11$\pm$0.06 & 20.1$\pm$0.3 & -0.72 & -0.07 \\
\hline
M31 & 0.7 & ... & Y & 3150 & 1.93 & 17.3 & ... & ... \\

\enddata
%% Text for table notes should follow after the \enddata but before
%% the \end{deluxetable}. Make sure there is at least one \tablenotemark
%% in the table for each \tablenotetext.
\tablecomments{Reading from the left, the columns are: Col. 1---name; col. 2---distance in
Mpc; col. 3---oxygen abundance estimated from
optical drift-scan spectroscopy by Hughes et al. (2011).
For each galaxy, one or more estimates of the
oxygen abundance have been made using different line
ratios and the calibrations described
in Kewley \& Ellison (2008), which use the O3N2 calibration from
Pettini \& Pagel (2004) as the base calibration. The
abundances given in the table are the means of these
estimates.
Col. 4---whether the
galaxy has PACS observations; col. 5---number of pixels
with S/N at 500 $\mu$m greater than 10; col. 6---mean value of $\beta$
for galaxy; col. 7---mean value of temperature for galaxy;
col. 8---Value of Spearman rank coefficient for relationship
between $\beta$ and galactocentric radius;
col. 9---Value of Spearman rank coefficient for relationship
between temperature and galactocentric radius.
}
\end{deluxetable*}

\section{The Calibration Sample and the Data}

Our main aim in this initial paper is 
not to provide a practical method for estimating the gas masses of
high-redshift galaxies, which would require a study
of a very large sample of galaxies covering a wide range of metallicity
and other parameters, but to investigate the ultimate potential of
this method. The biggest practical difficulty in
using this method is in measuring accurate dust temperatures.
Although the effect of an error in the dust temperature
on the estimate of the dust mass
is much less at the longest Herschel wavelength, 500 $\mu$m, than at the wavelengths
used by previous
space observatories, it is still approximately linear, a 10\% error in the
dust temperature leading to a $\simeq$10\% error in the dust mass.
There is an additional problem if there is a mixture of dust at different temperatures
within the telescope beam; the dust temperature obtained
by fitting a single-temperature modified blackbody
to this dust will then be systematically too high, leading to a systematic underestimate
of the dust mass (Eales, Wynn-Williams and Duncan 1989; Shetty et al. 2009a).
To overcome this problem as much as possible, we have restricted this
intial pilot study to galaxies that are well resolved by Herschel. There is a strong temperature
gradient in the dust in the Milky Way (Planck Collaboration 2011b) and we will
show in this paper that this is also true for most of the galaxies we
have studied. Therefore global submillimetre observations of a galaxy will almost certainly
contain dust with a range of temperatures, which will be a practical
problem in using this method to estimate the mass of the gas
in high-redshift galaxies detected by Herschel. However, by restricting our
study to galaxies which are large enough that we can investigate the distribution of
dust temperature within each galaxy, we have attempted to minimise this effect,
in order to explore the ultimate limits of this technique.

\begin{figure*}
\epsscale{.80}
%\plotone{Figure1.eps}
\centering
\includegraphics[trim=0.0mm 0.0mm 0.0mm 0.0mm,clip=true,width=0.68\textwidth]{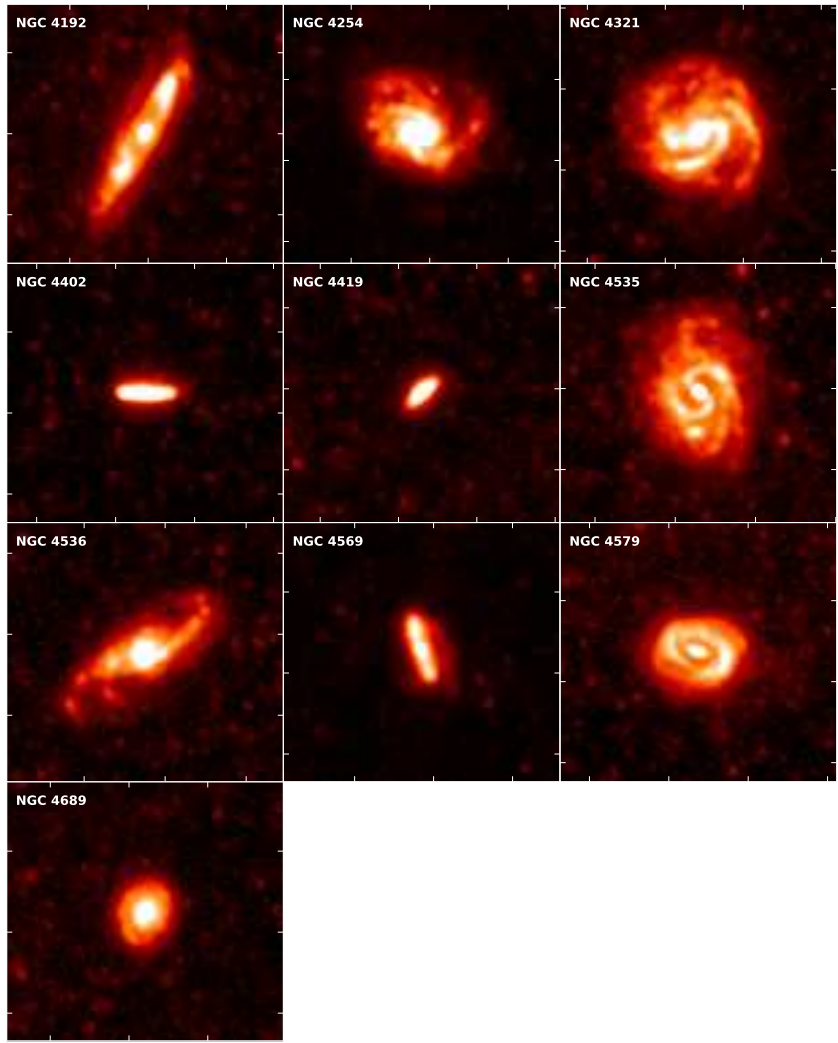}
\caption{Images at 250$\mu$m of the galaxies in the calibration sample. The tick
marks are at an interval of 3 arcmin.
\label{fig1}}
\end{figure*}

The sample is mostly drawn from two Herschel
key projects: the Herschel Reference Survey (HRS),
a survey with SPIRE at 250, 350 and 500 $\mu$m of
323 galaxies in a magnitude-limited and volume-limited sample of the local Universe
\citep{ale}, and the Herschel Virgo Infrared Cluster Survey (HeVICS, Davies
et al. 2010), 
a survey of the
Virgo Cluster at 110 and 160 $\mu$m with PACS and
at 250, 350 and 500 $\mu$m with SPIRE. Our method requires high-quality maps
made in the CO 1-0 line and in the 21-cm line, which cuts us down to 10
galaxies that have both been mapped in CO 1-0 at the Nobeyama Radio Observatory
\citep{kuno} and in the 21-cm line by the VLA as part of the VIVA survey (VLA
Imaging of Virgo in Atomic Gas - Chung et al. 2009). A requirement for our method
is that we need observations at $\lambda < 250\ \mu m$ to determine 
the dust temperature. Six of the galaxies are common to HRS and HeVICS and so
have PACS observations; for the others we have used archival Spitzer observations
at 70 $\mu$m, reprocessed as described by Bendo et al. (2011).
The reduction of the SPIRE and PACS data for the two key projects
is described by Smith et al. (2011a) and Davies et al. (2011).

Table 1 lists the galaxies and their basic properties. Figure 1
shows a montage of the 250-$\mu$m images of the galaxies.
Although this sample is not representative of
the entire galaxy population,
consisting of galaxies in the Virgo Cluster, 
we have partly 
mitigated
this problem by also considering two other galaxies:
our own (through the Planck observations) and the other
big spiral in the Local Group, M31. The analysis of 
the gas and dust in M31 is presented elsewhere
(Smith et al. 2011b) and we simply give the necessary results here. 

\begin{figure}
%\epsscale{.80}
%{Figure2.eps}
\centering
\includegraphics[trim=0mm 0mm 0mm 0mm,clip=true,width=0.34\textwidth]{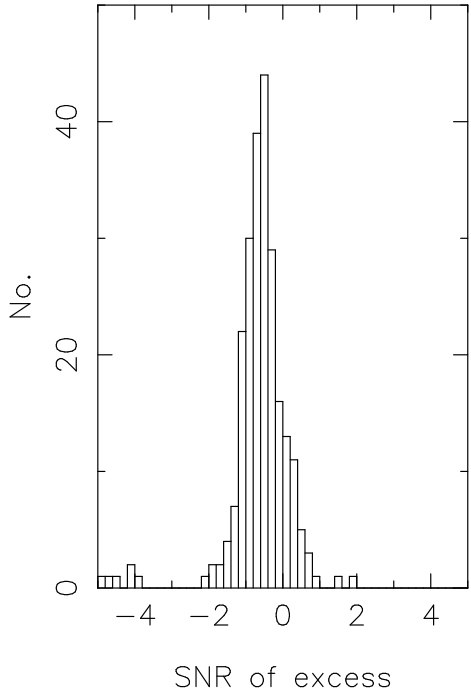}
\caption{Histogram for all pixels for which we fitted a modified blackbody
of
$(F_{500 \mu m} - F_{model})/\sigma$ in which $F_{500 \mu m}$ is the measured
flux at 500 $\mu$m, $F_{model}$ is the flux predicted by the single-temperature modified
blackbody, and $\sigma$ is the error (including the part of the calibration error
that is not correlated between bands but not the part that is
correlated between bands). There is no evidence for any excess emission at 500 $\mu$m.
\label{fig2}}
\end{figure}

\section{The Temperature of the Dust in the Galaxies}

We smoothed all the images to the resolution of the lowest-resolution
image, the 500 $\mu$m image, using the
method described in Bendo et al. (2010). We then regridded all the images
on to a pixel scale of 36 arcsec, which is the size of the 500-$\mu$m beam
(FWHM), ensuring that the data in each pixel is largely, although
not completely, independent.
We estimated the dust temperature by fitting single-temperature modified blackbodies 
($\rm F_{\nu} \propto B_{\nu}(T_d) \nu^{\beta}$) to the measured fluxes for each
pixel, allowing
the dust temperature ($\rm T_d$) and the dust emissivity index ($\beta$) to
vary independently and minimizing the
$\chi^2$ statistic. We restricted
all our analysis to pixels
in which the emission in each band was detected at $\rm > 5\sigma$.
Where PACS data existed, we fitted the modified blackbodies to
the 100-500 $\mu$m flux densities, and for the other galaxies
we used the Spitzer 70-$\mu$m image and the 250, 350 and 500 $\mu$m
measurements. We fitted the modified blackbody to the measured flux
densities
after convolving it with the appropriate filters
and after applying, for the SPIRE wavelengths, the `K4' correction
(SPIRE Observers' Manual). 
For the SPIRE measurements, we used the filter functions given for
extended sources in the SPIRE Observers' Manual. 
In calculating the $\chi^2$ statistic, we assumed 20\% calibration
uncertainties for PACS and Spitzer and two calibration errors
for the SPIRE measurements,
one of 5\% correlated over the three bands and one of 5\% uncorrelated
between the bands (SPIRE Observers' Manual).
We estimated the flux errors in each pixel by adding in quadrature
the calibration error
and the r.m.s. dispersion in the pixel values in an annulus around the galaxy.

\begin{figure}
%\epsscale{.80}
%\plotone{Figure3.eps}
\centering
\includegraphics[trim=0mm 0mm 0mm 0mm,clip=true,width=0.34\textwidth]{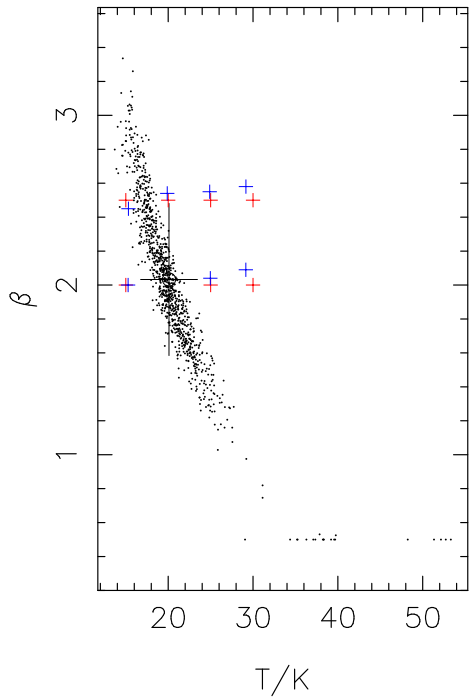}
\caption{The results of a Monte-Carlo simulation
in which we start with a single-temperature modified blackbody
with a 500-$\mu$m flux typical
of the pixels in the disk of M100, add typical noise, and then
use our fitting procedure to estimate the values of
$T_d$ and $\beta$. The black points show the results
for $\rm T_d=20 K$ and $\beta = 2$ with the large cross
showing the variance of the distribution along both axes.
The red crosses show the other combinations
of $T_d$ and $\beta$ we tried, and the blue crosses
show the mean values of the estimates from the fits.
\label{fig3}}
\end{figure}

For nine of the galaxies we found that the $\chi^2$ values of the fits
showed that
a single-temperature modified blackbody
is an adequate representation of the data. 
For M100 we found 20\% of the pixels had $\chi^2>2.71$, whereas we would have expected 10\%
by chance, so that for this galaxy there is evidence that the dust emission
from some pixels may not be adequately represented by a single-temperature modified
blackbody.
Submillimetre observations of some dwarf galaxies have found excess emission at
long wavelengths, which may indicate a large amount of very cold dust,
a change in the dust emissivity at long wavelengths, or for which there may
be some 
other explanation \citep{maud}. This excess has
not been seen in the HRS galaxies
(Boselli et al. 2010b). Nevertheless, we
looked for evidence for this in our sample by measuring
$(F_{500 \mu m} - F_{model})/\sigma$ for each pixel, in which $F_{500 \mu m}$ is
the flux density at 500 $\mu$m, $F_{model}$ is the flux at that wavelength given by
the best-fitting model, and $\sigma$ is the noise. In calculating $\sigma$ we have not 
included the part of the calibration error that is correlated between bands.
Fig. 2 shows a histogram of this quantity for all the pixels. There is no evidence
in these galaxies for any excess emission at 500 $\mu$m.

We estimated errors on each estimate
of temperature and $\beta$ using the method of Avni (1976).
Table 1 gives the 
weighted averages of the temperature and $\beta$ for each galaxy.
We have also included estimates of these quantities for M31
(Smith et al. 2011b), using the same technique as here.
With the exception of NGC 4536, the values of
$<T_d>$ are similar to the values
for the local interstellar dust estimated by COBE (17.5K - Boulanger et al.
1996)
and by Planck (17.9K - Planck Collaboration 2011c) and to the average value
for the dust in M31 (17.3 K - Smith et al. 2011b).
The similarity of the
values of $T_d$ and the
results of other recent Herschel studies (Bendo et al. 2011a, Foyle
et al. 2011; Smith et al. 2011b) suggest
that a very large fraction of the dust emission at $\rm \lambda > 100\ \mu m$
is from dust grains heated by the general interstellar radiation field rather than from 
dust
grains in the warmer environment of a star-formation region. 
We do not address here the question of whether this general interstellar radiation field
is dominated by the light from old stars (Bendo et al. 2011a) or whether it is
dominated by the light from the young OB stars (Foyle et al. 2011).
The values of $<\beta>$ are slightly but systematically higher than the
values for the Milky Way (1.8 - Planck Collaboration 2011b,c) and
for M31 (1.93 - Smith et al. 2011). We could possibly explain the difference
with the Planck results by some calibration issue between Planck and Herschel,
but the results for M31 have been obtained using similar data and method to that used
here. We do not have an explanation for the difference, but it is possibly
relevant that 
most of the gas in the Virgo galaxies is in the
molecular phase whereas most of the gas in M31 is in the atomic phase
although it is not obvious why this should produce
a change in $\beta$.

We made a rough estimate of possible systematic errors in the dust temperature
by repeating the analysis for the six galaxies with PACS observations but omitting
the measurements in the 100 $\mu$m band. One effect of fitting the model to
only four flux measurements (160-500 $\mu$m) is to increase the statistical
errors on the dust temperatures by a factor of $\simeq$2. There is no clear
direction to changes in the estimates of the dust temperature: for
three galaxies
the temperature estimate increases and for three it decreases. The
temperature change has a mean of $\delta T = 0.70K$ and a standard deviation
of 1.17 K, leading to a change in the estimate of the
mass of the ISM
with a mean of 6\% and a standard deviation of 12\%.
	 
We can investigate further that the dust emission is adequately represented by
dust at a single temperature by taking advantage of a well-known problem with
this kind of analysis.
If there is 
a range of dust temperature
along the line-of-sight for a particular galaxy, the 
values of temperature and $\beta$ produced
by fitting a modified blackbody will be systematically too high and too
low, respectively, which 
for an individual galaxy will lead to an underestimate of the
dust mass.
The important clue that this is
a significant problem within a sample of galaxies is
an inverse relationship between $\beta$
and temperature
\citep{Shetty1}.
Before we can look for this relationship, we need to model
another well-known problem: that the
fitting procedure itself can produce a spurious $\beta$-$T$ 
relationship \citep{Shetty2}. 
Figure 3 shows the results of
a 1000-run Monte-Carlo simulation in which we start with
a single-temperature modified blackbody with
a 500-$\mu$m flux typical
of one of the pixels in the disk of M100, add noise typical of
these pixels, and then use our fitting procedure to
estimate $\beta$ and $T$. As expected, the fitting
procedure produces 
a strong inverse correlation between $\beta$ and $T$.
However, the average value for both $\beta$ and
$T$ that is recovered from the simulation is always
very close to the input values, showing that there
is no systematic bias in the results of our fits.

\begin{figure}
%\epsscale{.80}
%\plotone{Figure4.eps}
\centering
\includegraphics[trim=0mm 0mm 0mm 0mm,clip=true,width=0.34\textwidth]{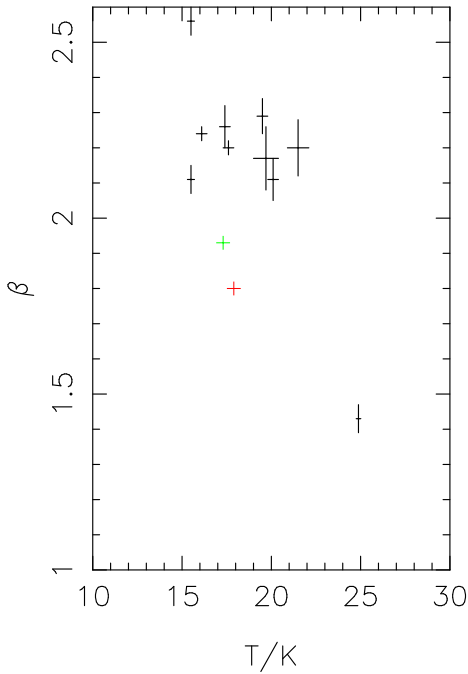}
\caption{The mean values of $T$ and $\beta$
for the galaxies in Table 1.
The crosses for these galaxies show the sizes of the 1$\sigma$ errors
on the means.
The red cross shows the estimated values of $T$ and $\beta$
from the Planck observations of high-latitude dust in the
Milky Way (Abergel et al. 2011) and the green cross shows
the average values of $T$ and $\beta$ for M31, estimated
using the same method we have employed here (Smith et al. 2011b). \label{fig4}}
\end{figure}

Figure 4 shows the measured value of
$<T>$ and $<\beta>$ plotted against each other
for
each galaxy. Although the errors on the average
values for the two quantities are still correlated,
the figure shows that the errors are much smaller than
the range of values for the two quantities.
Any inverse correlation between the
average quantities might then indicate the line-of-sight 
problem: the
existence of dust at a range of temperatures
in the galaxies with the ratio of warm-to-cold
dust varying between the galaxies. 
There is no strong relationship between the two, with
one
obvious outlier, the galaxy NGC 4536.
For NGC 4536, we used
a Spitzer 70-$\mu$m flux,
and
it seems likely that in this individual case
the low value of
$\beta$ and the high temperature are signs that
much of the 70-$\mu$m flux is from dust
that is warmer than the bulk of the dust.
Therefore, for this galaxy it seemed
likely that the estimate of the gas mass from the
dust emission would be too low, and
so
we decided to omit NGC 4536 from the calibration sample.
The other anomalous galaxy is M100
at the top of the diagram, but since there was no \`a priori reason to
believe that
a high value of $\beta$ will lead to an overestimate of the
dust mass and consequently
of the gas mass,
we retained this galaxy in the sample.

\begin{figure}
%\plotone{Figure5.eps}
\centering
\includegraphics[trim=0mm 0mm 0mm 0mm,clip=true,width=0.4\textwidth]{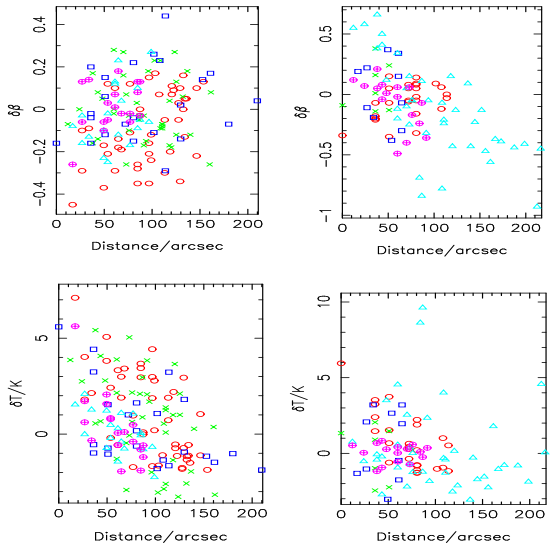}
\caption{Plots of $\beta$ (top) and temperature (bottom) versus
radius for each galaxy. For each galaxy, the quantity plotted
is the difference between the
value of $\beta$ or temperature in a pixel and the average value for
that galaxy.
The two panels on the left are for M100 (red circles), M99 (green
crosses), M98 (blue squares), M90 (light blue triangles) and
M58 (red crosses in circles). The two panels on the right
are for NGC 4535 (red circles), NGC 4419  (green
crosses), NGC 4402 (blue squares), NGC 4536 (light blue triangles) and
NGC 4689 (red crosses in circles).\label{fig5}}
\end{figure}

Figure 5 shows $T_d$ and $\beta$ plotted against the
distance from the centre of each galaxy, and
Table 1 gives the value of the Spearman correlation coefficient
for each correlation.
For eight of the remaining nine
galaxies, there is a 
trend for $T$ to increase
towards the centre of each galaxy (significant
in five cases),
which can be qualitatively explained by
the increase in the intensity of the
interstellar radiation field towards the centre
of a galaxy. This is similar to the radial gradient in
dust temperature measured by the Planck team for the
Milky Way (Planck Collaboration 2011b). The correlations between $\beta$
and galactocentric distance are generally weaker and
not in a consistent direction.
Since the errors on the estimates of $T$ and $\beta$ 
for individual pixels are large,
and because there is no compelling evidence for the variation of $\beta$ within
these galaxies,
in the rest of the analysis we assume the 
value of $\beta$ for each galaxy is a constant and 
use the value given in Table 1.

\section{Calibrating the Method}

The mass of hydrogen can 
be estimated from equation (2). It can also be estimated in the
standard way from
the CO and HI lines:

\smallskip

\begin{equation}
M_{H,meth\ 2} = M(HI) + ({ X \over 2 \times 10^{20}}) M(H_2) 
\end{equation}

\smallskip

\noindent in which $M(H_2)$ is the mass of molecular gas estimated
using an X factor of $\rm 2 \times 10^{20}\ cm^{-2} (K\ km\ s^{-1})^{-1}$
(Bolatto et al. 2008).

If we knew the value of the X-factor, we could estimate
$\eta_c$ for each galaxy by taking the ratio of the
masses estimated from the two equations. The mean value would
then give us the average calibration for the dust method,
with the variance giving us an estimate of the usefulness
of the method. Unfortunately, we don't, the uncertainty in the
X-factor being a perennial irritation in extragalactic
astronomy \citep{bell}.

We have tried to overcome this problem by 
making the hypothesis that there is a universal
value of X and $\eta_c$, at least for these nine
galaxies, and finding the minimum chi-squared
discrepancy from this hypothesis.
When estimating the gas mass from equation (2),
we used only pixels for which the 500-$\mu$m flux is 
detected at $>10\sigma$ and estimated a temperature for
each pixel by fitting a single-temperature modified blackbody
to the fluxes for that
pixel with the value of $\beta$ from Table 1.
We used the flux at 500 $\mu$m because the sensitivity of
this method to errors in dust temperature decreases with wavelength.
Our estimate of the gas mass for each galaxy is then the
sum of the values for the individual pixels.

Before making the 
alternative estimate of the gas mass from the CO/HI method, we convolved the
HI and CO maps to the same resolution as the Herschel images and put them
on the same pixel scale.
When estimating the gas mass from equation (3) we used the same pixels
that were used for estimating the gas mass from the dust method.
We estimated a total mass of atomic hydrogen for each galaxy
by summing the values in the HI map in these pixels
and a total mass of molecular hydrogen by summing the values
in the CO map for these pixels.

To apply the chi-squared method it is necessary to make some
assumption about the errors in the estimates
of
$M_{H}$ in
equation (2) and of $M(HI)$ and $M(H_2)$ in equation (3).
We assumed errors of 10\% in the estimates of
the total $M_{H}$ for each galaxy obtained
using the dust method, and we assumed the same error for the total
mass obtained from the HI image.
Based on our inspection of the CO maps, we concluded that
the errors in $M(H_2)$ are likely to be larger, and we conservatively
assumed that the errors in the total $M(H_2)$ for each galaxy are
30\%. We estimated the error in our estimate of
$\eta_c$ using the method of Avni (1976), assuming there
is one `useful' paraemter.

\begin{figure}
%\plotone{Figure6.eps}
\centering
\includegraphics[trim=0mm 0mm 0mm 0mm,clip=true,width=0.41\textwidth]{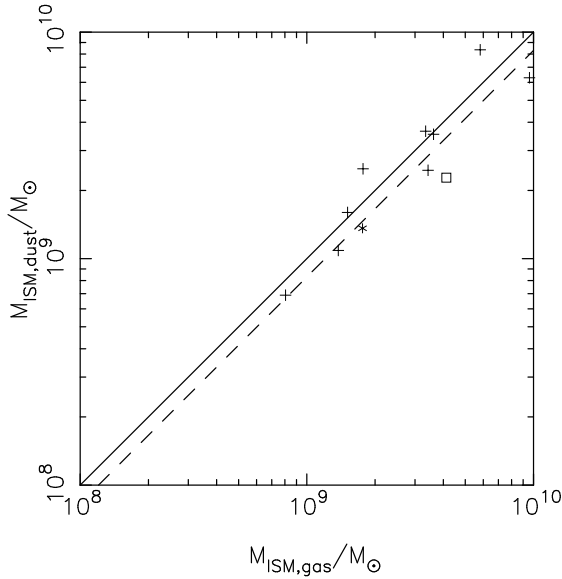}
\caption{The gas masses derived using the two alternative methods using
our best-fit values of the X factor and $\eta_c$.
The crosses show the values for the nine galaxies in our
calibration sample. The star shows the values for
M31 and the square shows the value for
NGC 4536. The solid line shows where the two gas masses are equal. The dashed
line shows the relationship predicted using the calibration from the Planck observations
of the Milky Way (Planck Collaboration 2011b).
\label{fig6}}
\end{figure}

We varied X between 0.1 and 10, finding the
best agreement with the model (minimum chi-squared)
with X=1.95 and $\eta_c = 1.21$, with an upper 1$\sigma$ limit
of 1.41 and a lower 1$\sigma$ limit of 1.07. 
If we omit the three galaxies without PACS measurements, we find
best agreement with the model with
X=2.93 and $\eta_c = 1.44$ with an upper
limit 1$\sigma$ limit of 1.71 and a lower 1$\sigma$ limit
of 1.28, slightly larger but consistent with the
value for the full sample.
Figure 6 shows the two estimates of the hydrogen mass plotted against
each other for all the galaxies.
We have included estimates
for NGC 4536, although we did not use it to estimate
$\eta_c$; as expected, the mass of hydrogen estimated from the dust emission
is systematically lower than for the other galaxies.
We have also included in the figure estimates of the
hydrogen mass for M31,
using the same procedure and the same values of $\eta_c$ and the X factor as
for the other galaxies.

Using the estimate for the full sample, we
can rewrite equation (2) in a simpler
way:

\smallskip

\begin{equation}
M_H = k {L_{\nu} \over B_{\nu}(T_d) }
\end{equation}

\smallskip
\noindent in which our estimate of $k$ 
is 640 kg m$^{-2}$ at 500 $\mu$m with an uncertainty of
approximately 20\%. We have not used our method
to estimate $\eta_c$ at 250 and 350 $\mu$m because of the
larger effect of temperature errors at these
wavelengths. However, we can make an approximate estimate
of the values of $k$ at these wavelengths by scaling
by $\nu^{-\beta}$, using the median value of $\beta$ in Table
1 (2.20). This gives values of $k$ at 250 and 350 $\mu$m
of 139 and 292 kg m$^{-2}$, respectively.

\section{Discussion}

We can obtain an upper limit
on the intrinsic error of our proposed method
of using the continuum dust emission to
estimate the gas mass
by calculating the dispersion
in Figure 6
of the ratio of the gas masses calculated using the two methods.
The standard deviation of
$log_{10} {M_{H, meth\ 1} \over M_{H, meth\ 2}}$ is
0.11, which gives an upper limit on the intrinsic error
in estimating the mass of the gas from the dust emission
of 30\%. This is an upper limit, because this error
term is also made up of observational errors and other 
possible astrophysical effects, such as variations in the
X factor from galaxy to galaxy and differing amounts of
CO-dark gas. However, even if we take
this upper limit as an actual estimate of the error in this
technique, the method is still clearly potentially useful
for estimating the mass of the gas in the hundreds of thousands
of high-redshift Herschel sources for which it will never be possible
to make CO observations.

There are several limitations in our work.
First, our sample does not represent very well
the entire galaxy population, since all the
galaxies are in the Virgo Cluster with metallicities
close to the solar value (Table 1).
The fact that both M31 and the Milky Way 
effectively have similar values of $\eta_c$ (Fig. 6) is
evidence that the cluster membership
is not leading to significant bias, but the metallicity issue
is an important one because
one would
expect
$\eta_c$ to have a dependence on metallicity
(James et al. 2002; Draine et al. 2007; Corbelli et al. 2011).
The relationship will be difficult to determine
by simply comparing the gas masses derived using the two methods, because
the X-factor is also likely to have a dependence on metallicity
(Wilson 1995; Arimoto et al.
1996; Boselli et al. 2002; Israel 2005;
Magrini et al. 2011. See Bolatto et al. [2008] for an alternative view), 
which also means of course
that
the metallicity issue does not give
the standard CO/HI method a clear advantage
for estimating the masses of the gas in high-redshift galaxies.
Nevertheless, despite the difficulty, for
turning this method into a practical one for estimating
the masses of the gas in high-redshift galaxies,
it will be important
to carry out a similar analysis to the one in this paper
but for a much larger sample of galaxies, 
with a wider range of properties, especially metallicity.

This is work for the future, but we note that a major difficulty
is that there are very few absolute measurements of the
metallicity in external galaxies. Most studies have
relied on relationships between line ratios and
metallicity that have been calibrated from absolute
measurements or theoretical models, and
uncertainties in these calibrations lead to an uncertainty
in the metallicity of a factor of $\simeq$4 (Moustakas et al. 2010; Hughes
et al. 2011; Magrini et al. 2011). We note in passing that the 
uncertainty in metallicity is a major problem with
estimating the mass-opacity coefficient of the dust using the method
of James et al. (2002) and Eales et al. (2010b), since this method relies
on accurate measurements of the metallicity in external galaxies
relative to the local ISM in our galaxy. Eales et al. (2010b) estimated
a mass-opacity coefficient from Herschel observations of M99 and M100,
finding a value much lower than predicted by dust models. Hughes
et al. (2011) have recently made new estimates 
of the metallicity
of the gas in M99 and M100 from existing data and find 
values four times greater than
the values used in Eales et al. (2010b), accounting for this discrepancy.

\begin{figure}
%\plotone{Figure7.eps}
\centering
\includegraphics[trim=0mm 0mm 0mm 0mm,clip=true,width=0.40\textwidth]{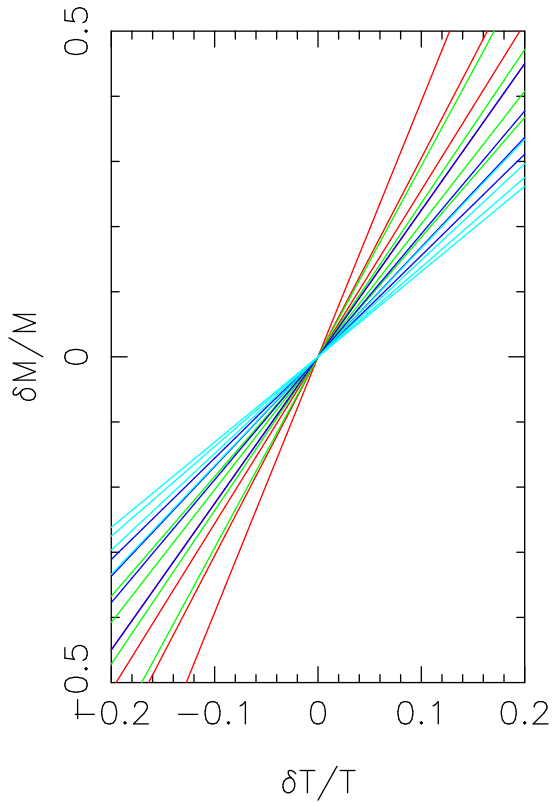}
\caption{Plot showing the effect of an error in the
estimate of dust temperature on the estimate of the
mass of dust in a galaxy. The different colours correspond
to different wavelengths: red---250 $\mu$m; green---350 $\mu$m;
dark blue---500 $\mu$m; light blue---850 $\mu$m. Curves are plotted
for four dust temperatures: 15K, 20K, 25K and 30K. For any particular
wavelength and fractional temperature error, the
effect on the estimate of the dust mass is greatest for the
coldest temperature.\label{fig7}}
\end{figure}

A second problem is that we have calibrated the dust method using
the standard CO/21-cm method, but if the latter
method is missing CO-dark gas, there will be a bias
in our calibration. This is not such a simple problem
to overcome because it requires observations of
this gas phase in nearby galaxies, which is 
difficult but can probably be done with
observations in the [CII] 158.9-$\mu$m line with Herschel
and SOFIA. 

We also emphasise that for this method to be effective it is
important to have accurate and unbiased estimates of the temperature
of the dust.
Fig. 7 shows how the
percentage error in gas mass
depends on the percentage error in the dust temperature for different
rest-frame wavelengths and dust temperatures. 
The effect of an error in dust temperature increases with decreasing
wavelength, and so in applying this method it is crucial
to measure the monochromatic luminosity at the longest possible
wavelength. For galaxies at $z > 1$, the effect of
an error in the dust temperature becomes very large even at the
longest possible Herschel wavelength. Therefore, for this method
to be practical at the highest redshifts, continuum measurements
from the ground will be crucial.

\begin{figure}
%\plotone{Fig8.eps}
\centering
\includegraphics[trim=0mm 0mm 0mm 0mm,clip=true,width=0.40\textwidth]{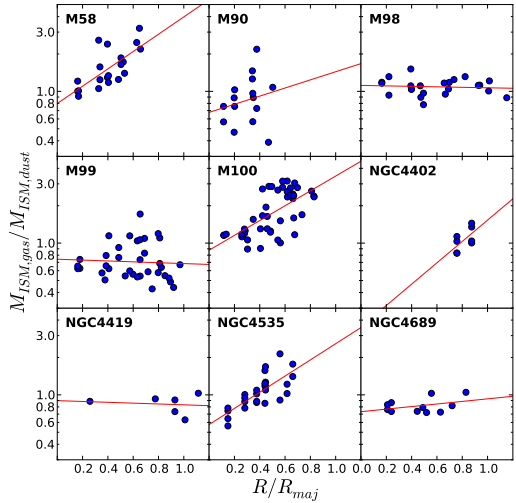}
\caption{Plots of the ratio of the two estimates of the gas mass
against the distance from the centre of each galaxy (see text for details).
\label{fig8}}
\end{figure}

Finally, we examine the potential of this method for estimating the
distribution of the ISM within an individual galaxy, since the Atacama Large Millimetre
Array will make it relatively easy to map the dust emission in a high-redshift
galaxy. Figure 8 shows the ratio of 
the two estimates of the gas mass for each pixel
plotted against distance from the
centre of each galaxy.
In making these estimates, we have used 
our best estimates of X and $\eta_c$ (see above) and
we have
corrected for the inclination of each
galaxy using the inclinations and position angles given in Chung et al. [2009]).
There are clear radial gradients that were effectively averaged over
when we carried out our global calibration.
These gradients
may be due to metallicity gradients, which
can effect both methods (Magrini et al. 2011), or might be due to other effects
such as the X factor for an individual galaxy being different from the
standard value. In order to use either method as a reliable way
of mapping the ISM in a high-redshift galaxy it will be crucial to
understand the cause of these gradients. This will be very tricky because
both the gas-to-dust factor and the X factor are likely to depend on metallicity
(Wilson 1995; Lisenfeld and Ferrara 1998; James et al.
2002; Boselli et al. 2002; Israel 2005; 
Magrini et al. 2011). 
We believe progress is most likely to be made by measuring
metallicity gradients for large numbers of galaxies with Herschel, HI and CO
observations. Fortunately, despite the uncertainty in the absolute values of
the metallicities of galaxies, most methods give similar results for the
metallicity gradients (Moustakas et al. 2010; Magrini et al. 2011).
Although the cause of these gradients is uncertain,
the plots show again that the method of using the dust emission to map
the ISM is a promising one. The r.m.s. dispersion of the log of the ratio
of the two gas estimates around the best-fit lines in Figure 8 is 0.118,
showing that, after a correction is made for the gradients,
the upper limit on the accuracy of one method (on the assumption that the
other method is correct) is 31\%, which is virtually the same as the accuracy forthe global estimates.
Since both methods rely on using tracers of the underlying gas (CO molecules
and dust grain) and since there are observational errors in both cases, we believe
the agreement between the two methods 
is remarkably good.
Therefore, if the obstacle of understanding the cause of the gradients can be
overcome, both methods should give consistent maps of the
ISM in high-redshift galaxies.

\acknowledgments

Herschel is an ESA space observatory with science instruments provided by 
European-led Principal Investigator consortia with important participation 
from NASA.
SPIRE has been developed by a consortium of institutes led by
Cardiff Univ. (UK) and including Univ. Lethbridge (Canada);
NAOC (China); CEA, LAM (France); IFSI, Univ. Padua (Italy);
IAC (Spain); Stockholm Observatory (Sweden); Imperial College
London, RAL, UCL-MSSL, UKATC, Univ. Sussex (UK); Caltech, JPL,
NHSC, Univ. Colorado (USA). This development has been supported
by national funding agencies: CSA (Canada); NAOC (China); CEA,
CNES, CNRS (France); ASI (Italy); MCINN (Spain); Stockholm
Observatory (Sweden); STFC (UK); and NASA (USA).


%% in parentheses, after the keyword, but they will not be verified.



\appendix

\section{The relation between submmillimetre luminosity and
gas mass}

The relationship obtained by the Planck team (Planck Collaboration 2011b) between
the optical depth of submillimetre emission at a frequency $\nu$
and the column-density of hydrogen (measured in atoms cm$^{-2}$) is
given by

\begin{equation}
\tau_{\nu}  = 1.1\times10^{-25} ({\lambda \over 250{\rm \mu m}})^{-1.8}
N_H
\end{equation}

\noindent We can translate this to a relationship between the submillimetre luminosity and
the mass of gas using Kirchoff's Law:

\begin{equation}
j_{\nu} = \kappa_{\nu} B_{\nu}(T)
\end{equation}

\noindent in which $j_{\nu}$ is the emissivity, $\kappa_{\nu}$ is the
absorption coefficient and $B_{\nu}(T)$ is the Planck function. 
We assume that the dust is at a single constant temperature and that the
dust is optically thin. The monochromatic submillimetre luminosity
is then given by

\begin{equation}
L_{\nu} = \int j_{\nu} dV = B_{\nu}(T) \int \kappa_{\nu} dV
\end{equation}

\noindent If $x$ is the distance along the line-of-sight,
the absorption coefficient is given by

\begin{equation}
\kappa_{\nu} = {d\tau_{\nu} \over dx} = 
1.1\times10^{-25} ({\lambda \over 250{\rm \mu m}})^{-1.8} n_H
\end{equation}

\noindent in which $n_H$ is now the number of hydrogen atoms per cubic
centimetre. By substituting A4 in A3, we obtain the relationship:

\begin{equation}
L_{\nu} = B_{\nu}(T) 1.1\times10^{-25} ({\lambda \over 250{\rm \mu m}})^{-1.8} \int n_H dV
\end{equation}

\noindent From which we obtain:

\begin{equation}
M_{H} = {1.52 \times 10^2 \times L_{\nu} \over
B_{\nu}(T_d) \times ({\lambda \over 250})^{-1.8} }
\end{equation}

\noindent in which we have converted everything into S.I. units except
wavelength
($\lambda$), which
is measured in microns.
Luminosity is measured in Watts Hz$^{-1}$ sr$^{-1}$.
In this derivation we have assumed that the dust is precisely tracing the
dust and has the same gas-to-dust ratio as in the Milky Way. If that is not
the case, the constant of proportionality in equation A6 will simply scale
with the gas-to-dust ratio.

\end{document}